# A Convoloutional Neural Network model based on Neutrosophy for Noisy Speech Recognition

Elyas Rashno
Computer Engineering Department
Iran University of Science and Technology
Tehran,Iran
elyas.rashno@gmail.com

Ahmad Akbari
Computer Engineering Department
Iran University of Science and Technology
Tehran,Iran
akbari@iust.ac.ir

Babak Nasersharif
Computer Engineering Department line
K.N.Toosi University of Technology
Tehran,Iran
bnasersharif@kntu.ac.ir

*Abstract*— Convolutional neural networks are sensitive to unknown noisy condition in the test phase and so their performance degrades for the noisy data classification task including noisy speech recognition. In this research, a new convolutional neural network (CNN) model with data uncertainty handling; referred as NCNN (Neutrosophic Convolutional Neural Network); is proposed for classification task. Here, speech signals are used as input data and their noise is modeled as uncertainty. In this task, using speech spectrogram, a definition of uncertainty is proposed in neutrosophic (NS) domain. Uncertainty is computed for each Time-frequency point of speech spectrogram as like a pixel. Therefore, uncertainty matrix with the same size of spectrogram is created in NS domain. In the next step, a two parallel paths CNN classification model is proposed. Speech spectrogram is used as input of the first path and uncertainty matrix for the second path. The outputs of two paths are combined to compute the final output of the classifier. To show the effectiveness of the proposed method, it has been compared with conventional CNN on the isolated words of Aurora2 dataset. The proposed method achieves the average accuracy of 85.96 in noisy train data. It is more robust against noises with accuracies 90, 88 and 81 in test sets A, B and C, respectively. Results show that the proposed method outperforms conventional CNN with the improvement of 6, 5 and ۲ percentage in test set A, test set B and test sets C, respectively. It means that the proposed method is more robust against noisy data and handle these data effectively.

## I. INTRODUCTION

The uncertainty of each element is not considered and described in classical fuzzy set. The traditional fuzzy set describes the membership degree with a real number $\mu_A(x) \in [0,1]$ [1]. In this situation, if $\mu_A(x)$ is uncertain, it is not defined by the crisp value [2]. In some applications, such as information fusion, expert systems and belief systems, we should consider not only the truth-membership supported by the evidence, but we need to consider the falsity-membership and uncertainty-membership in these applications. It is hard for classical fuzzy set to solve these problems as well [2]. Therefore, it is required to introduce the model that can handle the indeterminacy. Neutrosophic (NS) set attempts to solve this problem with considering uncertainty that it quantified explicitly the truth-membership, indeterminacy-membership and falsity-membership. This assumption is very important in many applications [3].

Neutrosophy is a branch of philosophy which studies the nature and scope of the neutralities and their interactions which is the basis of NS logic [4]. NS theory was first proposed by Smarandache in 1995 [5, 6]. This theory was applied for image processing first by Guo et. al [3] and then it has been successfully used for other image processing domains including image segmentation[3,7-10, 28 and 35], image thresholding[11], image edge detection[12], retinal image analysis[13-21], content-based image retrieval [22 and 23], liver image analysis[24 and 25], breast ultrasound image analysis [26], data classification [27], uncertainty handling [24 and 36], image and data clustering [29 and 36-37].

Traditionally, neural networks are used for prediction and classification purposes when they are adapted with a number of input values. However, any prediction or classification may be associated with a degree of uncertainty. In the neural network community, there are two kind of uncertainty associated with neural network outcomes: uncertainty in the training dataset and uncertainty in the model structure [30 and 31]. A convolutional neural network (CNN) is a class of deep neural networks, most commonly applied to analyzing visual imagery. CNNs use a variation of multilayer perceptron's designed to require minimal preprocessing. They are also known as shift invariant or space invariant artificial neural networks (SIANN), based on their shared-weights architecture and translation invariance characteristics [31].

CNN's have also been widely used for speech recognition task as feature extractor and also classification models. For example, CNN and DBN have been applied to Large-Scale ASR task as feature extractors [32]. In [33], very deep CNNs (up to ten layers) is used for robust speech recognition. In another research [34], CNN has been used as a robust feature extractor from noisy speech spectrogram in two ways: fixed resolution and multiresolution convolution filters.

The main contribution of this work is to model uncertainty of speech data in NS domain so that noisy speech data can be handled efficiently by CNN classification models. In the first step, we consider speech spectrogram as a image. This image is transferred to NS domain and then a new definition of data uncertainty is proposed for spectrogram pixels. Uncertainty matrix with the same size of spectrogram is the result of this task. Then, a tow parallel paths CNN is proposed. One path is used for spectrogram and the other path is used for uncertainty matrix. Two paths classification results are combined with different combination methods.

The rest of this paper is organized as follows: Section II present review on NS. The proposed method is discussed in Sections III. Experimental results are reported in section IV. Finally, this work is concluded in Section V.

## II. REVIEW ON NEUTROSOPHIC SET

NS is a powerful framework of Neutrosophy in which neutrosophic operations are defined from a technical point of view. In fact, for each application, neutrosophic sets are defined as well as neutrosophic operations corresponding to that application. Generally, in neutrosophic set A, each member x in A is denoted by three real subsets true, false and indeterminacy in interval [0, 1] referred as T, F and I, respectively. Each element is expressed as x(t, i, f) which means that it is t% true, i% indeterminacy, and f% false. In each application, domain experts propose the concept behind true, false and indeterminacy [3].

To use NS in image processing domain, the image should be transferred into the neutrosophic domain. Although the original method for this transformation was presented by Guo et all. [2], these methods completely depends on the image processing application. An image g has a dimension of M×N. g can be shown with three subsets: T, I and F in NS domain. So, pixel p(i,j) in g is shown with PNS(i, j) = {T(i, j), I(i, j), F(i, j)}) or PNS (t, i, f) . T, I and F indicate white, noise and black pixel sets, respectively. PNS (t, i, f) provides useful information about white, noisy and black percentage in this pixel that is %t to be a white pixel, %i to be a noisy pixel and %f to be a black pixel. T, I and F are computed as follows [2-3].

$$T(i,j) = \frac{\overline{g(i,j)} - \overline{g}_{min}}{\overline{g}_{max} - \overline{g}_{min}} \quad (1)$$

$$F(i,j) = 1 - T(i,j) \quad (2)$$

$$I(i,j) = \frac{\delta(i,j) - \delta_{min}}{\delta_{max} - \delta_{min}} \quad (3)$$

$$\overline{g(i,j)} = \frac{1}{w^2} \sum_{m=-\frac{w}{2}}^{m=\frac{w}{2}} \sum_{n=-\frac{w}{2}}^{n=\frac{w}{2}} g(i+m, j+n) \quad (4)$$

$$\delta(i,j) = \left| g(i,j) - \overline{g(i,j)} \right| \quad (5)$$

where $g$ is gray scale image, $\overline{g}$ is filtered image g with average filter, w is window size for average filter, $\overline{g}_{max}$ and $\overline{g}_{min}$ are the maximum and minimum of the $\overline{g}$, respectively, $\delta$ is the absolute difference between g and $\overline{g}$, $\delta_{max}$ and $\delta_{min}$ are the maximum and minimum values of δ, respectively[3].

## III. PROPOSED METHOD

The main motivation for this research is that how noisy speech data can be handled efficiently so that CNN based Automatic Speech Recognition (ASR) systems can be made more robust against noise. Here, noise is interpreted as indeterminacy. Indeterminacy for any data type can be modeled in NS domain. Therefore, the main contributions of this research can be summarized in two main steps. First, indeterminacy for speech data is proposed in NS domain. Then, a novel CNN model is proposed in which data indeterminacy is considered.

### A. Indeterminacy of speech data

Speech signals are transferred to spectrogram domain and then the appeared noise in spectrogram is interpreted and handled in NS domain. In NS domain, a filter is considered with the length of f and t in spectral and temporal domains, respectively. The proposed method for indeterminacy computation is presented in Eqs (6)-(10):

$$T(i,j) = \frac{\overline{g(i,j)}}{g_{mean}} \quad (6)$$

$$I(i,j) = \frac{\delta(i,j)}{\delta_{mean}} \quad (7)$$

$$\overline{g(i,j)} = \frac{1}{t \times f} \sum_{m=-\frac{t}{2}}^{m=\frac{t}{2}} \sum_{n=-\frac{f}{2}}^{n=\frac{f}{2}} g(i+m, j+n) \quad (8)$$

$$\delta(i,j) = \left| g(i,j) - \overline{g(i,j)} \right| \quad (9)$$

where $\overline{g}_{mean}$ and $\delta_{mean}$ represent the average of all data points in $\overline{g(i,j)}$ and $\delta(i,j)$, respectively. A rectangular filter in Eq. (8) is applied to compute $\overline{g}$ matrix. It means that speech spectrogram is blurred with a rectangular filter to compute $\overline{g}$. Then, the difference between $g$ and $\overline{g}$ is considered as $\delta$. Indeterminacy is achieved by dividing $\delta$ over $\delta_{mean}$. This idea can be interpreted by this fact that the bigger difference between a data point in spectrogram and its neighbors tends to the bigger indeterminacy. Indeterminacy set for noisy and clean spectrograms have been shown in Fig. 1. In spectorgram, noise is spread along frequence axis, therfore, rectangular filter is considered in NS domain to cover more information in this axis. The spectrogram of a clean speech and its corresponding indeterminacy set are shown in Fig 1.(a) and Fig 1.(b), respectively. Also, these sets for a noisy speech are shown in Fig 1.(c) and Fig 1.(d).

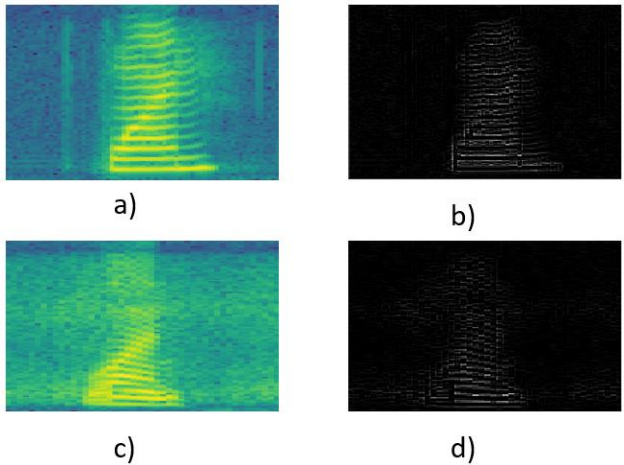

Fig. 1. Indeterminacy set: (a): clean speech Spectrogram, (b): Indeterminacy of (a), (c): noisy speech Spectrogram for a, (d): Indeterminacy of (c),

### B. CNN model with indeterminacy

The proposed model for considering indeterminacy in CNN include parallel two paths CNN depicted in Fig. 2. After experiments on several models, the best performance was achieved by the following structure including 5 convolutional layers, 3 pooling layers and 3 fully-connected layers.

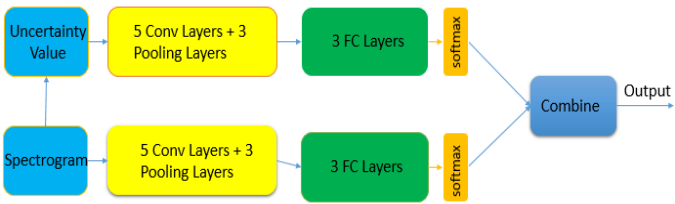

Fig. 2. Proposed CNN structure with considering indeterminacy

The spectrogram of speech signals is input of a CNN with 5 convolution layers, 3 pooling layers followed by 3 fully connected layers. The indeterminacy of spectrogram is also input of another CNN with the same structure. The results of these networks are combined to make the output. More details of the proposed CNN structure is demonstrated in Table I.

## IV. EXPERIMENTAL RESULTS

### A. Dateset

We evaluate the proposed method using isolated words of the Aurora2 dataset. The number of isolated words in Aurora 2 is equal to 11. Aurora2 is divided into train and test sets. Train set is further divided into clean and noisy sets. Test set is also divided into 3 sets A, B and C, each one contains clean and noisy sets. Test set A includes Subway, Car, Babble, Exhibition additive noises. In test set B, Restaurant, Street, Airport and Train Station noises are included. Finally, in test set C, Subway and Street noises are used where the channel is different from the channel used for recording set A and B. Each type of noise is considered with different signal to noise ratios including: (SNRs) -5, 0, 5, 10, 15 and 20 db.

### B. Parameter setting

The proposed CNN structure has been implemented with Tensorflow libraries in python. All activation functions in output layers are considered as LERU while in convolution layers tangent hyperbolic is used. Stochastic gradient decent with the batch size of 32 is used to train the CNN model. The maximum number of iterations is set to 3000. The structure of the CNN layers, layer sizes, number of filters in each layer, size of each filter, stride and output sizes are reported in table 1.

TABLE I. THE PROPOSED STRUCTURE OF CNN LAYERS.

| Layer type | Input size | No. filter | filter size | Stride | Output |
|---|---|---|---|---|---|
| *Convolutional* | 925×1475×3 | 64 | 5×5×3 | 5×5 | 185×295×64 |
| *Pooling* | 185×295×64 | --- | 3×3 | 2×2 | 93×147×64 |
| *Convolutional* | 93×147×64 | 64 | 5×5×64 | 1×1 | 93×147×64 |
| *Pooling* | 93×147×64 | --- | 3×3 | 2×2 | 48×74×64 |
| *Convolutional* | 48×74×64 | 128 | 3×3×64 | 1×1 | 48×74×128 |
| *Convolutional* | 48×74×128 | 128 | 3×3×128 | 1×1 | 48×74×128 |
| *Convolutional* | 48×74×128 | 238 | 3×3×128 | 1×1 | 48×74×128 |
| *Pooling* | 48×74×128 | --- | 5×3 | 5×6 | 10×13×128 |
| *FC* | 10×13×128 | --- | --- | --- | 384 |
| *FC* | 384 | --- | --- | --- | 192 |
| *FC* | 192 | --- | --- | --- | 11 |

### C. Results

The structure for network layers in table 1 is used as a basic model referred as CNN. CNN is used in two parallel paths, one path with the input of spectrogram and another path with the

Input of indeterminacy. This proposed model is called as NCNN. Based on train and test sets in Aurora2 dataset, two models are trained with clean and noisy data. Table 2 reports the accuracy of ASR system with CNN and NCNN classifiers, which are trained by clean train data. When the network is trained with clean data, it is not robust enough against noisy test data. The accuracies of CNN for noisy test sets A, B and C are 63.50, 61.25 and 46.77, respectively. The proposed NCNN structure, improves the accuracies for noisy test sets A, B and C to 74.71, 68.52 and 62.76, respectively. It means that NCNN is more robust against noisy test sets with the average accuracy of 68.66 in comparison with CNN with 57.17. The proposed method outperforms the basic CNN with 9. 7 and 16 percent for test sets A, B and C, respectively.

TABLE II. CNN AND NCNN WITH CLEAN TRAINING SET

|  | ASR accuracy | | | |
|---|---|---|---|---|
|  | Test Set A | Test Set B | Test Set C | Average |
| CNN | 63.50 | 61.25 | 46.77 | 57.17 |
| NCNN | 74.71 | 68.52 | 62.76 | 68.66 |

The next experiment it to train the models with noisy data. Table 3 reports the accuracy of CNN and NCNN for 3 test sets. In this case, NCNN achieves 6, 5, and 2 percent of better accuracy.

TABLE III. CNN AND NCNN WITH NOISY TRAINING SET

|  | ASR accuracy | | | |
|---|---|---|---|---|
|  | Test Set A | Test Set B | Test Set C | Average |
| CNN | 83.93 | 81.31 | 79.23 | 81.49 |
| NCNN | 89.45 | 86.96 | 81.47 | 85.96 |

One of the most important parameter in network is window size in NS domain. Window size determines how data are transferred into NS domain. Bigger window size leads to transfer more general spectrogram data to NS domain. On the other side, smaller window size keeps local data in NS domain. Therefore, small window size models local indeterminacy while large window size models global indeterminacy. To show the effect of window size, an experiment has been done. In this experiment, NCNN has been configured with 4 window sizes 20×40, 30×10, 10×30 and 30×30. Table 4 reports the accuracy of NCNN with clean train data for different window sizes. It can be concluded that the best performance for test sets A and B is achieved with window size 10×30 while for test set C, window size 30×10 has a better accuracy. Therefore, window size 10×30 leads to the highest average accuracy of 68.66 percent for all test sets.

TBALE IV. DIFFERENT WINDOW SIZES FOR CLEAN TRAIN DATA.

| Window size | ASR accuracy | | | |
|---|---|---|---|---|
|  | Test Set A | Test Set B | Test Set C | Average |
| 20 × 40 | ٧١,٣٤ | ٦٤,٣٧ | ٥٨,٥٩ | ٦٤,٧٧ |
| 30 × 10 | ٧٣,٨٦ | ٦٤,٩٨ | ٦٣,٥١ | ٦٧,٤٥ |
| 10 × 30 | ٧٤,٧١ | ٦٨,٥٢ | ٦٢,٧٦ | ٦٨,٦٦ |
| 30 × 30 | ٧٠,٤٩ | ٦٠,١٩ | ٥٤,٠١ | ٦١,٥٦ |

The same experiment has been done to evaluate the effect of window size in NCNN with noisy train data. In this case, large window size is the best choice and leads to the best performance in all test sets. The reason is that when spectrogram contains noise, signal frequencies are dispread in along time axis. Therefore, bigger window size in NS domain can save more information of signal and then better model the spread noise. Table 5 reports the accuracy of NCNN with noisy train data for 4 windows sizes.

TABLE V. DIFFERENT WINDOW SIZES FOR NOISY TRAIN DATA.

| Window size | ASR accuracy | | | |
|---|---|---|---|---|
|  | Test set A | Test Set B | Test Set C | Average |
| 20 × 40 | 89.45 | 86.96 | 81.47 | 85.96 |
| 30 × 10 | 88.78 | 85.78 | 80.86 | 85.14 |
| 10 × 30 | 87.92 | 86.49 | 79.80 | 84.74 |
| 30 × 30 | 88.75 | 85.7 | 79.23 | 84.56 |

TBALE VI. DIFFERENT COMBINATIONS METHODS IN NCNN WITH CLEAN TRAIN DATA.

| Combination method | ASR Accuracy | | | |
|---|---|---|---|---|
|  | Test Set A | Test Set B | Test Set C | Average |
| Product | 89.45 | 86.96 | 81.47 | 85.96 |
| Sum | ٨٨,٧٨ | ٨٥,٧٨ | ٨٠,٨٦ | ٨٥,١٤ |
| Maximum | 87.35 | 85.79 | ٧٩,٥٠ | 84.21 |

The last issue in the proposed method is that, how two paths from spectrogram and indeterminacy can be combined. There are several options for networks combination. In this research, 3 methods including sum, product and maximum have been investigated. The accuracies of 3 combination methods in NCNN with clean and noisy train data are reported in tables 6 and 7, respectively. As it is clear from reported results, combination with product achieves the best accuracy for ASR system in 3 test sets A, B and C.

TBALE VII. DIFFERENT COMBINATIONS METHODS IN NCNN WITH NOISY TRAIN DATA.

| Combination method | ASR Accuracy | | | |
|---|---|---|---|---|
|  | Test Set A | Test Set B | Test Set C | Average |
| Product | ٧٤,٧١ | ٦٨,٥٢ | ٦٢,٧٦ | ٦٨,٦٦ |
| Sum | ٧٣,٥٤ | ٦٥,٧٣ | ٦٢,٥٤ | ٦٧,٢٧ |
| Maximum | ٧٣,٢٥ | ٦٦,٥٧ | ٦١,١٢ | ٦٦,٩٨ |

The proposed scheme for ASR has been evaluated in noisy signals with 8 types of noise including Babble, Car, Exhibition, Subway, Airport, Restaurant, Street and TrainStation. These noises were applied to train and test sets A, B and C with 6 SNRs -5, 0, 5, 10, 15 and 20. The average accuracy of the proposed method applied on test sets A, B and C with 10 noise types have been illustrated in Fig. 3. Note that set A includes Babble, Car, Exhibition and Subway noises, test B includes Airport, Restaurant, Street and TrainStation and test set C contains Street and Subway noises. Each bar in Fig. 3 represents the average accuracy of each noise type in all SNRs. It can be concluded that the proposed method is more robust comparison with conventional CNN specially in cases such as Car, Airport and Subway noises with accuracies 90, 88 and 81 in test sets A, B and C, respectively. Finally, the average accuracies of the proposed method in each SNR are depicted in Fig. 4.

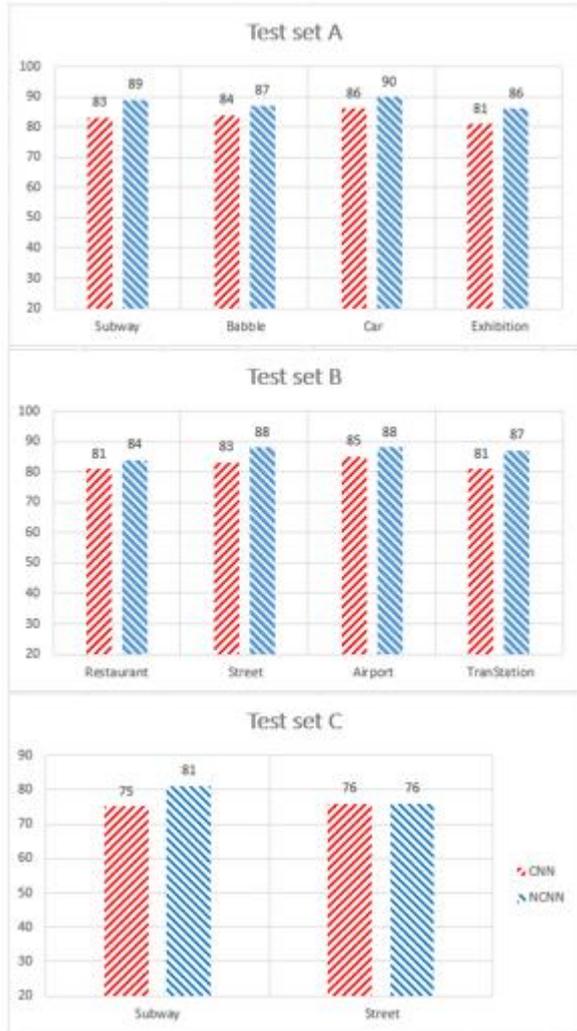

Fig. 3. Average recognition accuracy over all SNR values separated for different noise types and three test sets ( A , B , C)

V. CONCLUSION AND FUTURE WORKS

In this work, a two parallel paths CNN classification model with data uncertainty handling was proposed for classification task. Speech signals were used as input data and their noise was modeled as uncertainty in NS domain.

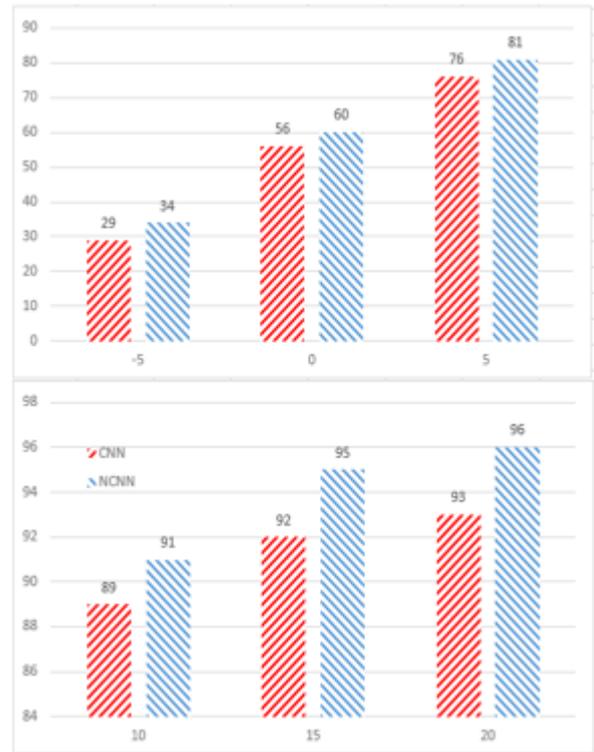

Fig. 4. Average word error rate over all noise type and test sets which are separated for SNR values.

The proposed method had benefits in noisy data handling. Experimental results showed that when the classification model is trained with clean data, it is robust against noisy data in the test phase. Also, when the classification model is trained with noisy data, it is more robust against noisy data in test phase. This behavior was concluded with comparison with convention CNN. Future efforts will be directed towards using the proposed model in other applications such as image classification by proposing uncertainty for image pixels in NS domain. Finally, using the proposed uncertainty model in other deep learning networks such as LSTM can be considered as future works.